\documentclass[10pt,twocolumn,letterpaper]{article}

\usepackage{iccv}
\usepackage{times}
\usepackage{epsfig}
\usepackage{graphicx}
\usepackage{amsmath}
\usepackage{amssymb}
\usepackage{caption}
\usepackage{cite}
\usepackage{xcolor}
\usepackage{subfig}
\usepackage{xspace}

\usepackage[breaklinks=true,bookmarks=false]{hyperref}
\hypersetup{
  colorlinks
}
\iccvfinalcopy 

\def\iccvPaperID{48} 

\begin{document}

\title{Studying the Effects of Self-Attention for Medical Image Analysis}

\author{Adrit Rao$^{1}$\quad Jongchan Park$^{2}$\quad Sanghyun Woo$^{3}$\quad Joon-Young Lee$^{4}$\quad  Oliver Aalami$^{5}$\\
		\normalsize $^{1}$Palo Alto High School \qquad
		\normalsize $^{2}$Lunit Inc. \qquad
		\normalsize $^{3}$KAIST  \qquad
		\normalsize $^{4}$Adobe Research  \qquad
		\normalsize $^{5}$Stanford University\\
		{\footnotesize \tt adrit.rao@gmail.com \quad jcpark@lunit.io \quad shwoo93@kaist.ac.kr \quad	jolee@adobe.com \quad aalami@stanford.edu}}

\maketitle


\begin{abstract}
When the trained physician interprets medical images, they understand the clinical importance of visual features. By applying cognitive attention, they apply greater focus onto clinically relevant regions while disregarding unnecessary features. The use of computer vision to automate the classification of medical images is widely studied. However, the standard convolutional neural network (CNN) does not necessarily employ subconscious feature relevancy evaluation techniques similar to the trained medical specialist and evaluates features more generally. Self-attention mechanisms enable CNNs to focus more on semantically important regions or aggregated relevant context with long-range dependencies. By using attention, medical image analysis systems can potentially become more robust by focusing on more important clinical feature regions. In this paper, we provide a comprehensive comparison of various state-of-the-art self-attention mechanisms across multiple medical image analysis tasks. Through both quantitative and qualitative evaluations along with a clinical user-centric survey study, we aim to provide a deeper understanding of the effects of self-attention in medical computer vision tasks.
\end{abstract}

\vspace{-3mm}
\section{Introduction}\label{sec:Intro}

The ability to leverage deep learning and computer vision-based techniques and methods for the automated, accurate, robust, and interpretable classification of medical images has been widely studied \cite{shen2017deep, litjens2017survey, lee2017deep}. By doing so robustly, we can potentially increase diagnostic accuracy and increase screening efficiency and productivity \cite{ker2017deep}. When developing computer vision-based systems to aid physicians, it is important to design the underlying task to be as similar as possible to the medical specialists. Additionally, when deploying these systems, interpretability is critical for clinical decision-making. Therein lies the value in making computer vision systems perform computation similar to human cognition. However, computational systems do not necessarily perform similar to humans in terms of visual cognition and it is important to integrate this capability.

\begin{figure}[h]
    \centering
    \includegraphics[width=\linewidth]{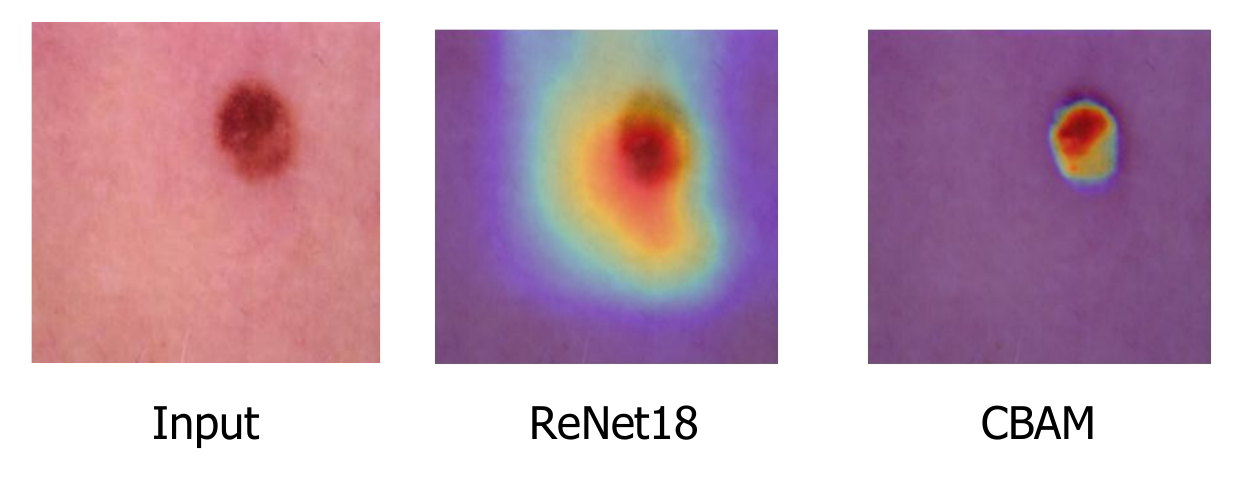}
    \caption{\textbf{Heat-map comparison between standard ResNet-18 and ResNet-18 + CBAM} on a benign skin image. The minimal addition of the visual attention mechanism (\hyperref[fig:CBAM_OVERVIEW]{CBAM}~\cite{woo2018cbam}) in the residual block (Sec.~\hyperref[sec:model_arch]{4.2}) enabled more focus onto the significant pigmented portion of the image making the final classification represent a region of more clinical importance while disregarding other features.}
    \label{fig:skin-cancer}
\end{figure}

\noindent\paragraph{Cognitive Visual Attention} One of the main cognitive capabilities which the trained domain specialist leverages is attention or ``focus'' \cite{chun2001visual}. Attention is the ability by which the human brain processes visual information while also evaluating the relevance of input features. This cognitive process allows for selective concentration onto a discrete stimulus while disregarding other perceivable stimuli \cite{allport1989visual}. As a general example, we can take a hypothetical scenario in which a person is viewing a landscape. Within this landscape is a certain object of interest. In order to effectively classify the object, the human brain applies this method of visual attention in order to use the majority of brain capacity to carry out the process on a certain region of interest (ROI). From a medical standpoint, in a chest CXR interpretation scenario, for example, the medical specialist will likely use attention subconsciously to automatically evaluate the clinical relevance of visual features with the advantage of also knowing the presenting symptoms of the patient and indication for the imaging study. With this, they are able to disregard unnecessary features (background, noise, etc.) and make a final diagnostic decision based solely on more relevant factors (abnormal lesions, opacity, etc.) \cite{bertram2016eye}. This leads to a more robust final diagnosis. Replicating this capability within medical computer vision solutions is very important. 

\noindent\paragraph{Visual Attention Mechanisms} However, in comparison to the cognitive capability of attention, the standard and widely used convolutional neural network (CNN) classifier analyzes features more generally and is not guaranteed to extract relevant clinical information similar to the trained domain specialists subconscious \cite{albawi2017understanding}. The ability to artificially replicate or mimic this capability within neural networks can enable similar evaluation of clinical feature relevance in images which can potentially lead to more robustness in analysis and increased accuracy. A notable innovation in the computer vision field is the self-attention mechanism~\cite{woo2018cbam,wang2017residual,wang2018non,cao2019gcnet,huang2019ccnet,hu2018squeeze}. Without any explicit supervision, these attention mechanisms learn to focus on important feature values in a data-driven manner. Self-attention mechanisms are trained end-to-end together with the original CNN backbone architectures without any changes in the training process. Thus, using self-attention mechanisms within the standard CNN can have many benefits on medical visual recognition tasks in terms of accuracy, interpretability, and robustness. Fig.~\ref{fig:skin-cancer} shows a visualization of activation heat-maps (Grad-CAM \cite{selvaraju2017grad}) from a standard CNN (ResNet-18 \cite{he2016deep}) and an attention-augmented CNN (ResNet-18 + CBAM \cite{woo2018cbam}) on a skin cancer dataset sample. ResNet-18 is a widely used CNN architecture employed for many image classification tasks. The attention-augmented ResNet-18 has the sole modification of added visual attention mechanisms (\hyperref[fig:CBAM_OVERVIEW]{CBAM}) subsequently following convolution before the skip connection in the residual blocks. Notice how the addition of attention enables the model to focus fully on the important mole region while the standard CNN is less attentive to the mole and has more of a distributed attention.

This study aims to understand the value of self-attention mechanisms within standard computer vision models for approaching the ability to mimic the medical specialist's ability to evaluate the importance of features and their clinical relevancy. By doing so, we aim to understand the value of attention and how it can enable models to focus on more important clinical features. We present an experimental setting in which we use the standard ResNet-18
backbone~\cite{he2016deep} for performing multiple experiments across medical imaging datasets (multiple data modalities) by augmenting the residual blocks to accommodate 3 state-of-the art attention mechanisms as per original placement (CBAM~\cite{woo2018cbam}, SE~\cite{hu2018squeeze}, GC~\cite{cao2019gcnet}). Each of these mechanisms performs differently in terms of computation. We perform validation through a standard quantitative accuracy measure (average AUC-ROC across 3 tests) and qualitative heat-map visualizations. We notice significant increases in accuracy and through a visualization, we also observe significant changes in heat-map activation over more clinically relevant features. To further understand the benefits from a clinical standpoint, we have expert medical specialists (dermatologists and radiologists) interpret the visualizations and provide insight into which models and attention mechanisms focus on the most relevant clinical features and lesions.
\vspace{-1.5mm}
\section{Related Work}

\subsection{Medical Image Classification} 

\textit{Computer vision aided medical diagnosis} is an important topic in the research community \cite{shen2017deep, litjens2017survey, lee2017deep}. The ability to automate medical imaging diagnosis at the point-of-care in a robust fashion can lead to more accurate and objective clinical evaluation as well as increased screening efficiency and quality control \cite{esteva2021deep}. Applications of computer vision include chest x-ray \cite{rajpurkar2017chexnet}, CT scan \cite{bhandary2020deep, grewal2018radnet, lakshmanaprabu2019optimal}, MRI classification \cite{korolev2017residual, liu2018applications} and many more. Several systems can accurately classify images at near medical specialist precision. However, most vision-based standard CNN analysis methods do not necessarily employ similar evaluation techniques as the trained specialist. Specialists are trained over many years to understand the clinical relevance of features. Thus, while interpreting images, they are able to disregard features that are not important in making a targeted diagnosis. The ability to develop vision systems that are coherent with this capability is important. The main cognitive capability which the trained specialist employs is attention. Attention enables the ability to evaluate the importance (clinically) of input visual information and perform classification through analysis of a certain region of interest. Similarly, state-of-the-art self-attention mechanisms are able to mimic this capability to a certain extent computationally. Thus, there could be potential value in the integration of visual self-attention within medical computer vision systems to approach specialist evaluation, increase accuracy, and all around the robustness of clinical prediction systems.

\subsection{Attention Mechanisms} 


Following the human visual perception process~\cite{attention1,attention2,attention3}, the most intuitive way of modeling visual attention is to relatively scale up important information (i.e. pay attention) and scale down less important information. The initial works~\cite{wang2017residual,hu2018squeeze,woo2018cbam} of attention in visual recognition use attention maps to dynamically scale intermediate feature values in CNNs.
One of the pioneering approaches is the Residual Attention Network (RAN)~\cite{wang2017residual}. RAN uses an additional attention branch with downsampling convolutions and upsampling layers to generate the attention mask which is the same size of the intermediate feature map. This direct computation is simple and intuitive and improves baseline performance yet the computational cost is quite high. The Squeeze-and-Excitation (SE)~\cite{hu2018squeeze} network is also a prominent approach that focuses on channel attention. For each given intermediate feature map, an SE mechanism generates a per-channel attention value from the global-average-pooled features. SE has been shown to improve performance with minimal overhead~\cite{hu2018squeeze}.
The Style-based Recalibration Module (SRM)~\cite{lee2019srm} is a simple yet powerful channel attention module that accounts for channel statistics (mean and standard deviation) when scaling the channel values. After pooling the statistics, SRM uses a channel-wise fully connected (CFC) layer where each channel's attention weights are computed by a linear combination of the two statistics. The CFC layer in SRM is extremely efficient making both the computational and parametric overhead minimal.
The Convolutional Block Attention Module (CBAM)~\cite{woo2018cbam} is a computationally efficient method that decomposes the heavy attention generation into separate dimensions. Specifically, while RAN directly generates full-sized attention maps, CBAM generates 2D spatial and 1D channel attention maps. CBAM has been shown to improve performance in various tasks consistently and reduce the overall computational overhead~\cite{woo2018cbam}.
In the NLP field, the majority of self-attention mechanisms use attention maps to utilize the long-range dependencies among semantic tokens. Recent self-attention mechanisms in visual recognition models~\cite{wang2018non,huang2019ccnet,cao2019gcnet} are also equipped with such long-range dependencies. 
Non-local Neural Networks (NL)~\cite{wang2018non} is the first piece of work in the visual recognition field to model the long-range dependencies among spatial locations. Most of the previous methods computed attention with limited context, however, NL~\cite{wang2018non} uses the attention map to softly aggregate the information for all the points in the feature map. That is, all the relevant information in the entire feature map can be added to each individual point in the feature map. However, one of the drawbacks of NL~\cite{wang2018non} is the high computational cost, and to resolve this, CCNet~\cite{huang2019ccnet} proposed to approximate the full attention process into separate cross-shaped processes. Aggregating two cross-shaped attention can effectively approximate the effect of NL~\cite{wang2018non}. Following CCNet~\cite{huang2019ccnet}, GCNet~\cite{cao2019gcnet} also solved the high-computation issue of NL~\cite{wang2018non} by simplifying the NL block and inheriting the bottleneck structure of SE~\cite{hu2018squeeze}. Through experimentation, GCNet achieves superior performance in comparison to NL in object detection, segmentation, classification, and action recognition tasks.

The objective of this study is to empirically verify the effectiveness of state-of-the-art self-attention mechanisms in various medical image analysis tasks. With attention mechanisms, neural networks can potentially start to approach the capability of understanding the clinical importance of features. We choose 3 self-attention mechanisms to compare: SE~\cite{hu2018squeeze}, CBAM~\cite{woo2018cbam}, and GC~\cite{cao2019gcnet}. SE is chosen because it is one of the most widely used self-attention mechanisms with the main focus on channel scale re-calibration, CBAM is chosen because it considers both channel and spatial dimensions, and GC is chosen because it is one of the strongest method with the non-local long-range dependency modeling. In the following sections, we will cover each method in detail and the results derived through validation.

\section{Attention Mechanisms in Detail}\label{sec:Attention}

\subsection{General Formulation}
Most of the self-attention mechanisms, including the three we compare in this study, are self-contained. \textit{Self-contained} means that the inputs and the outputs of the mechanism are the same allowing them to be integrated at any location in a CNN architecture without modifying other parts. Any self-contained self-attention mechanism would fit into the following equation shown below (Eq.~\ref{eq:eq1}):
\begin{equation}\label{eq:eq1}
\begin{split}
    F^{l+1}&=SA(F^l) \\
    F^l &\in \mathbb{R}^{C\times H\times W} \\
    F^{l+1} &\in \mathbb{R}^{C\times H\times W}
\end{split}
\end{equation}
\noindent where $F$ indicates the intermediate feature of a typical 2D CNN, $l$ indicates the current layer index, and \{$C, H, W$\} indicate the size of channel, height, and width respectively.

The self-attention allows the model to discover the most important task-relevant feature points. In practice, we dynamically compute the attention map of the feature map using its pooled~\cite{hu2018squeeze,woo2018cbam,cao2019gcnet} or raw features~\cite{wang2018non}.
At each layer, SE~\cite{hu2018squeeze} uses the global-average-pooled feature as the statistics of each input and computes the scale factor for each feature channel, CBAM~\cite{woo2018cbam} uses a similar manner with both global average and global maximum statistics in both channel and spatial dimensions and GC~\cite{cao2019gcnet} computes a softmax attention map to aggregate the global statistics over all the spatial locations and then computes a context feature to be added to the input feature map. Technical details for each of the self-attention mechanisms used in this study will be elaborated in the following sub-sections (\ref{sec:se}, \ref{sec:cbam}, \ref{sec:gc}).


\subsection{Squeeze-and-Excitation~\cite{hu2018squeeze}}\label{sec:se}

\begin{figure}[h!]
    \centering
    \includegraphics[width=\linewidth]{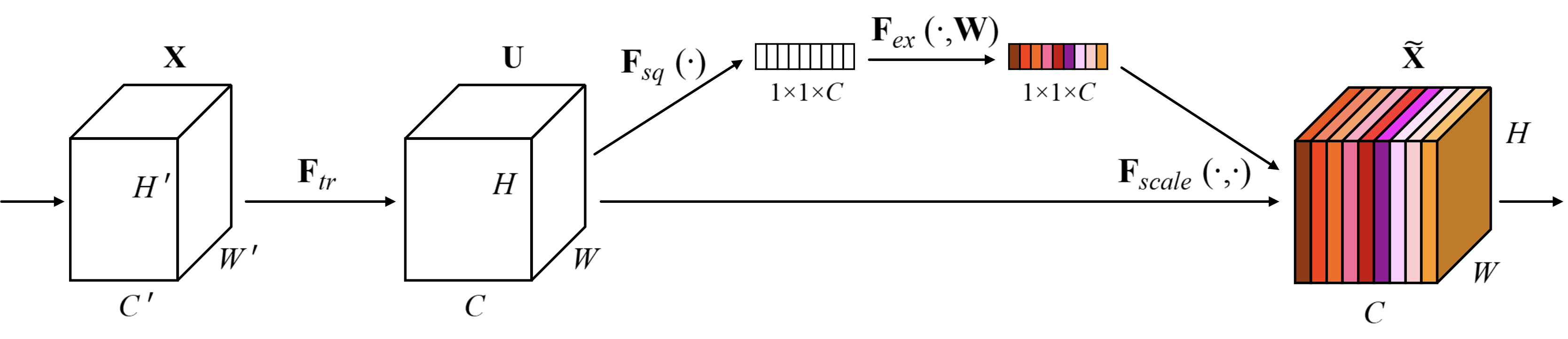}
    \caption{\textbf{Squeeze-and-Excitation (SE) network}. Architecture diagram taken from the original paper~\cite{hu2018squeeze}}
    \label{fig:SE_OVERVIEW}
\end{figure}

Hu \textit{et al}.~\cite{hu2018squeeze} proposed Squeeze-and-Excitation (SE), a self-attention mechanism focused on the channel dimension. SE is referred to as a `channel recalibration' method. That is, each channel's magnitude can be explicitly tuned according to the values in other channels. As illustrated in Fig.~\ref{fig:SE_OVERVIEW}, the first step of SE is to gather the global information in each channel through global average pooling (GAP). The term `global' indicates that the spatial dimensions (height and width) are reduced and the method only utilizes one pooled value for each channel. The second step of SE is to use the globally-pooled feature vector and two consecutive fully-connected (dense) layers with one ReLU layer in between. The intermediate channel size is $\frac{1}{16}$ of the input channel size and last channel size is equal to the input channel size. The last step of the SE block is to apply sigmoid to the last feature vector and multiply it back to the original full feature map. Eq.~\ref{eq:SE1} contains the described steps:
\begin{equation}\label{eq:SE1}
\begin{split}
    f_{ch}&=GAP(F)\\
    a_{ch}&=\sigma(FC_{\frac{c}{r}\rightarrow c}(ReLU(FC_{c\rightarrow\frac{c}{r}}(f_{ch}))))\\
    SA(F)&=F*a_{ch}
\end{split}
\end{equation}
\noindent where $\sigma$ indicates the sigmoid function, $GAP$ indicates the global average pooling function, $r$ indicates the reduction ratio for the intermediate channel, and $FC$ indicates fully-connected layers with input/output channels specified.
Similar to Eq.~\ref{eq:eq1}, $F\in\mathbb{R}^{C\times H\times W}$, $f_{ch}\in\mathbb{R}^{C\times 1\times 1}$, and $a_{ch}\in\mathbb{R}^{C\times 1\times 1}$. The final multiplication shown in Eq.~\ref{eq:SE1} is broadcasted along the spatial dimensions.


\subsection{Convolutional Block Attention Module~\cite{woo2018cbam}}\label{sec:cbam}

\begin{figure}[h!]
    \centering
    \includegraphics[width=\linewidth]{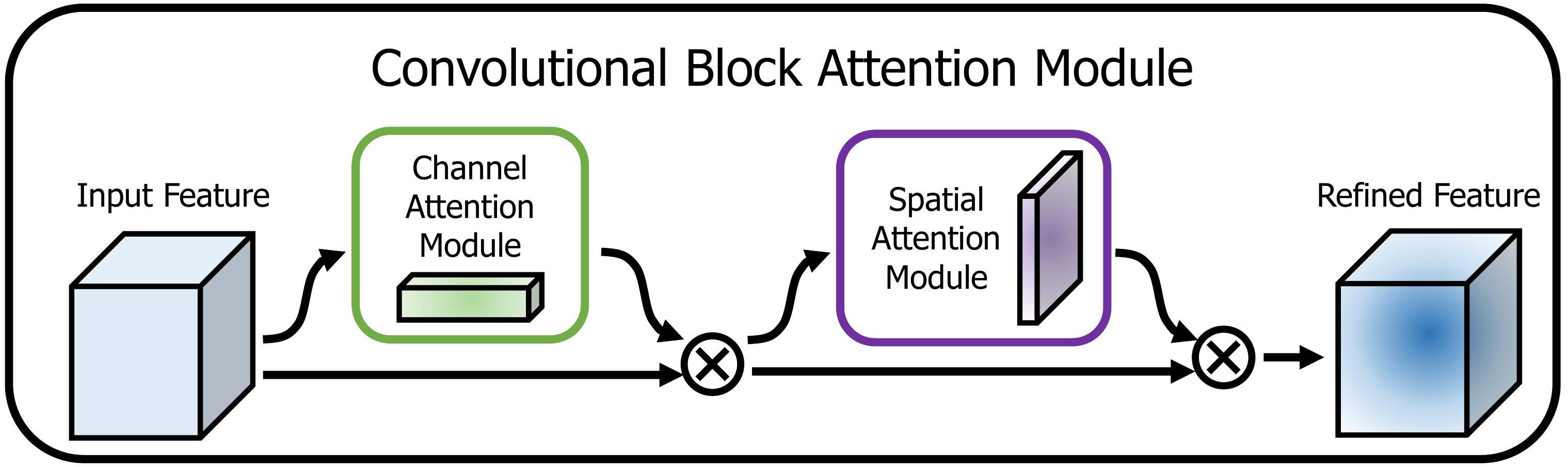}
    \caption{\textbf{Overall Convolutional Block Attention Module (CBAM)}. Diagram taken from the original paper~\cite{woo2018cbam}.}
    \label{fig:CBAM_OVERVIEW}
\end{figure}

Woo \textit{et al}.~\cite{woo2018cbam} proposed Convolutional Block Attention Module (CBAM), a self-attention mechanism designed to make use of both the channel and spatial dimensions. The direct computation of a 3D attention tensor is quite heavy, similar to RAN~\cite{wang2017residual}, roughly doubling the overall computation. CBAM decomposes the 3D attention tensor into 1D channel attention and 2D spatial attention and applies them sequentially to the input feature in order to reduce the computational overhead of the attention mechanism. The design is illustrated in Fig.~\ref{fig:CBAM_OVERVIEW}. SE utilizes the global average pooled statistics to calculate the attention weights while CBAM utilizes two statistics: global average and global maximum. The two statistics are experimentally shown to be complementary as using both statistics is better than using a single statistic. The channel and spatial attentions are sequentially applied to the input feature map $F$ as shown in Eq.~\ref{eq:CBAM_OVERVIEW}:
\begin{equation}\label{eq:CBAM_OVERVIEW}
\begin{split}
    SA(F)&=SA_{sp}(SA_{ch}(F)).
\end{split}
\end{equation}
where $SA_{ch}$ denotes the channel attention sub-module and $SA_{sp}$ denotes the spatial attention sub-module which follows after the channel attention sub-module computation. 

\begin{figure}[h!]
    \centering
    \includegraphics[width=\linewidth]{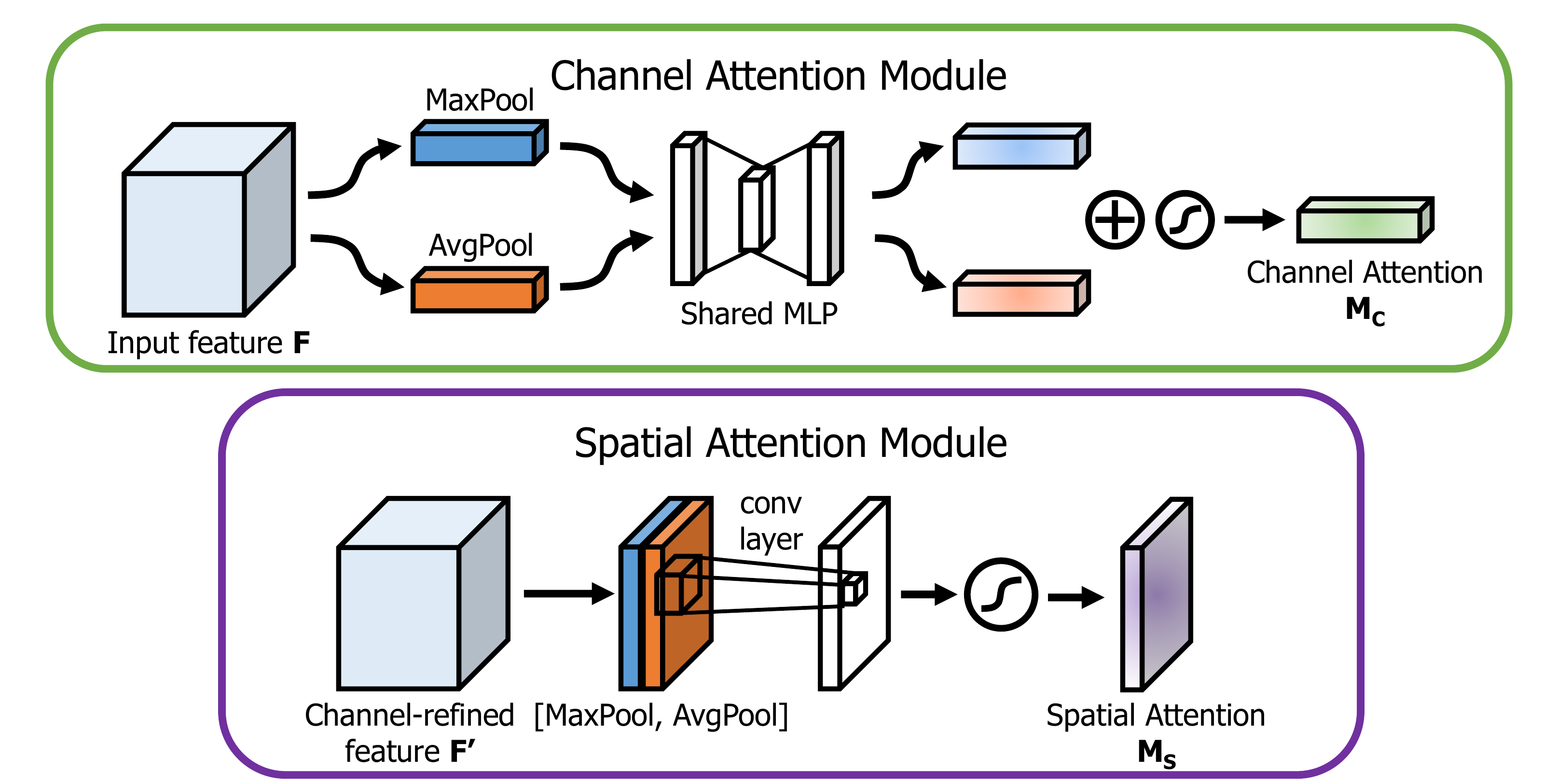}
    \caption{\textbf{Channel and spatial attention sub-modules}. Diagram taken from the original paper~\cite{woo2018cbam}.}
    \label{fig:spatial-and-channel-module}
\end{figure}

\noindent\paragraph{Channel attention module} The structure of the channel attention module is illustrated in Fig.~\ref{fig:spatial-and-channel-module} (top). The attention tensor for the channel dimension is a 1D tensor. To efficiently calculate the 1D tensor, the global average and the global maximum values along each channel are pooled. After this, the 1D features are fed into a 2-layer multi-layered perceptron with a sigmoid layer at the end. The mathematical notation of the channel attention calculation is:
\begin{equation}\label{eq:ch-attention-module}
\begin{split}
    f_{ch-avg}&=GAP(F)\\
    f_{ch-max}&=GMP(F)\\
    SA_{ch}(F)&=\sigma(MLP(f_{ch-avg})+MLP(f_{ch-max}))*F
\end{split}
\end{equation}
where \(\sigma\) denotes the sigmoid function, \textit{MLP} is the 2-layer multi-layered perceptron with two fully-connected layers and one ReLU layer in between, $f_{ch-avg}$ and $f_{ch-max}$ are global average pooled / global max pooled features along the channel dimension, where $f_{ch-avg}$ and $f_{ch-max}$\(\in \mathbb{R}^{C\times 1\times 1}\). The final output of the channel attention module is the original 3D CNN feature multiplied by the 1D attention tensor with broadcasting along the spatial dimension.

\noindent\paragraph{Spatial attention module} The structure of the spatial attention module is illustrated in Fig.~\ref{fig:spatial-and-channel-module} (bottom). The architecture follows the same structure as the channel attention module the only difference being the fact that the spatial attention module focuses on the spatial dimension. The mathematical notation for the spatial attention calculation is:
\begin{equation}\label{eq:sp-attention-module}
\begin{split}
    F_{ch}&=SA_{ch}(F)\\
    f_{sp-avg}&=GAP_{sp}(F_{ch})\\
    f_{sp-max}&=GMP_{sp}(F_{ch})\\
    SA_{sp}(F_{ch})&=\sigma(Conv_{7\times 7}([f_{sp-avg}, f_{sp-max}])*F_{ch}.
\end{split}
\end{equation}
\noindent Note that the input to the spatial attention module is the output from the channel attention module, $F_{ch}$. As written in Eq.~\ref{eq:sp-attention-module}, the spatially avg/max pooled feature $f_{sp-avg}$ $f_{sp-max}$\(\in \mathbb{R}^{1\times H\times W}\) are fed into a convolutional layer to compute the spatial attention tensor $M_s$\(\in \mathbb{R}^{1\times H\times W}\).


\subsection{Global Context Network~\cite{cao2019gcnet}}\label{sec:gc}

\begin{figure}[h!]
    \centering
    \includegraphics[width=0.4\linewidth]{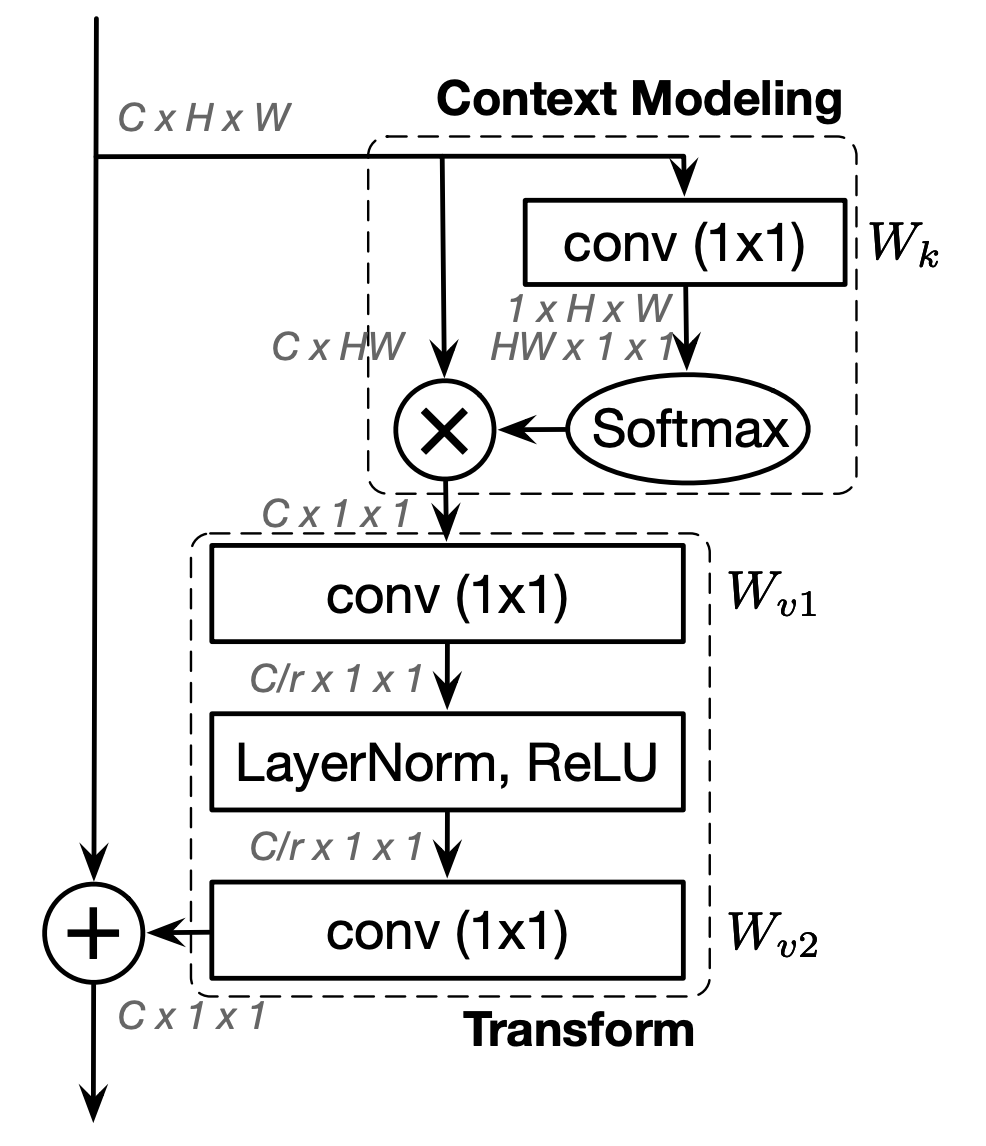}
    \caption{\textbf{Global Context Block architecture}. Diagram taken from the original paper~\cite{cao2019gcnet}}.
    \label{fig:gcnet}
\end{figure}

Cao \textit{et al}.~\cite{cao2019gcnet} proposed GCNet (Global Context Network), an efficient yet powerful self-attention mechanism with context aggregation. Prior to GCNet, Wang \textit{et al}. proposed Non-Local Neural Networks (NL)~\cite{wang2018non} for modeling contextual information for long-range dependencies. NL uses the key-query-value architecture of typical memory architectures and models the dense key-query affinity matrix between all pixel pairs in the feature map. All the values are aggregated using the affinity matrix thus, all the relevant information from all pixel locations are added to each pixel in the feature map. However, the dense affinity matrix modeling requires heavy computations and a large number of parameters to train. GCNet simplifies the architecture of NL and further reduces computation with the bottleneck architecture of SE~\cite{hu2018squeeze}. The architecture is illustrated in Fig.~\ref{fig:gcnet}. 
The forward path of GC is composed of the following steps:
\begin{equation}\label{eq:gcnet}
\begin{split}
    SA_{gc}(F)=SA_{tfm}(SA_{agg}(F))+F
\end{split}
\end{equation}
\noindent where $agg$ indicates context aggregation, $tfm$ indicates feature transform for context information. As shown in Eq.~\ref{eq:gcnet}, GCNet computes a context feature to be added back to the input feature map $F$. $ctx$ stands for `context'. As shown in Eq.~\ref{eq:gcnet_detail}, the context computation is composed of context aggregation and context transformation:
\begin{equation}\label{eq:gcnet_detail}
\begin{split}
    SA_{agg}(F)&=\text{Softmax}(\text{Conv}(F))*F\\
    SA_{tfm}(F)&=\underset{\frac{c}{r}\rightarrow c}{\text{Conv}}(\text{LN}(\underset{{c\rightarrow \frac{c}{r}}}{\text{Conv}}(F))).\\
\end{split}
\end{equation}
\noindent The first convolution in the context aggregation stage generates a softmax map to aggregate the feature values at each pixel locations. The resulting context vector is transformed with $SA_{tfm}$ which has the bottleneck structure as SE~\cite{hu2018squeeze}. The dense relationship modeling in NL~\cite{wang2018non} is reduced to a simple convolution layer and a softmax layer. After the context aggregation, a single global feature vector is computed and is transformed before adding back to the original input feature. One downside would be the lack of the location-specific context aggregation due to the simplification and all spatial locations will have the same context added back.

Throughout the paper, we strictly follow the original design and the hyper-parameters of each attention mechanism. However, we do not use the same backbone as the original papers. As all the attention mechanisms are of self-contained self-attention, they can be placed at any point within the backbone. Thus, to keep placement consistency, we position all mechanisms at the same location within a standard CNN backbone. Details are elaborated in Sec.~\ref{sec:model_arch}.



\section{Methods}

In this section, we describe and cover the methodology used to develop our experimental setting environment. The goal of our study is to understand the value of self-attention mechanisms and the potential performance increases and robustness which it provides for medical computer vision systems. We first cover the different medical datasets we use (\ref{sec:datasets}) and then go over the model architectures and attention placement (\ref{sec:model_arch}) with implementation details (\ref{sec:implemen}). 

\subsection{Datasets}\label{sec:datasets}

We perform our study across 4 medical image datasets. This is done to provide a more robust and comprehensive comparison of self-attention across different data modalities (skin, CXR, MRI, CT). For training, we use a standard 80\% training and 20\% validation split. We perform a quantitative statistical evaluation across all datasets and a qualitative user study across the skin cancer and CXR datasets.

\noindent\paragraph{Skin Dataset} The Skin Cancer Dataset consists of 3,297 processed skin images of mole lesions split into malignant (disease) and benign (normal) classes. The dataset was originally collected by the The International Skin Imaging Collaboration (ISIC) organization \cite{isicarchive} and made open-source. Differentiating factors between classes are mainly visual feature differences in the pigmented mole lesions \cite{jerant2000early}. 


\noindent\paragraph{CXR Dataset} The CXR (chest radiograph) dataset consists of 5,863 chest x-ray (anterior and posterior) images of normal and pneumonia classes from the open-source Chest X-Ray Images for Classification repository (UCSD) \cite{kermany2018labeled}. Differentiating factors between image classes include hazy shadowing in an pneumonia labeled CXR image \cite{cozzi2020chest}.


\noindent\paragraph{MRI Dataset} The MRI (magnetic resonance imaging) image dataset consists of 3,264 images of the human brain split into the classes of tumorous and no tumor from an open-source repository on Kaggle \cite{sartaj_2020}. The main visual differentiating factor between the classes are the tumorous lesions which are typically circular and in a different shade compared to the other parts of the brain MRI \cite{bhattacharyya2011brain}.


\noindent\paragraph{CT Dataset}

The CT (computed tomography) dataset consists of 812 CT scan images spanning the classes of COVID-19 positive and negative. The dataset is from the open-source UCSD COVID-CT repository \cite{yang2020covid}. The main visual differentiating factor between the classes are the ground-glass opacity, vascular enlargement and white/hazy shadowing within a COVID-19 positive CT scan \cite{he2020sample}.

\subsection{Model Architectures}\label{sec:model_arch}

Across all experiments, we use the ResNet-18~\cite{he2016deep} architecture as the backbone. ResNet-18 acts as as good backbone architecture due to it being widely used in a multitude of classification tasks. Following the PyTorch implementation of ResNet-18 for ImageNet\footnote{\url{https://github.com/pytorch/vision/blob/master/torchvision/models/resnet.py}}, we have made a minor modification on the final pooling layer and use a global average pooling instead of the fixed sized average pooling.

\begin{figure}[h!]
    \centering
    \includegraphics[width=\linewidth]{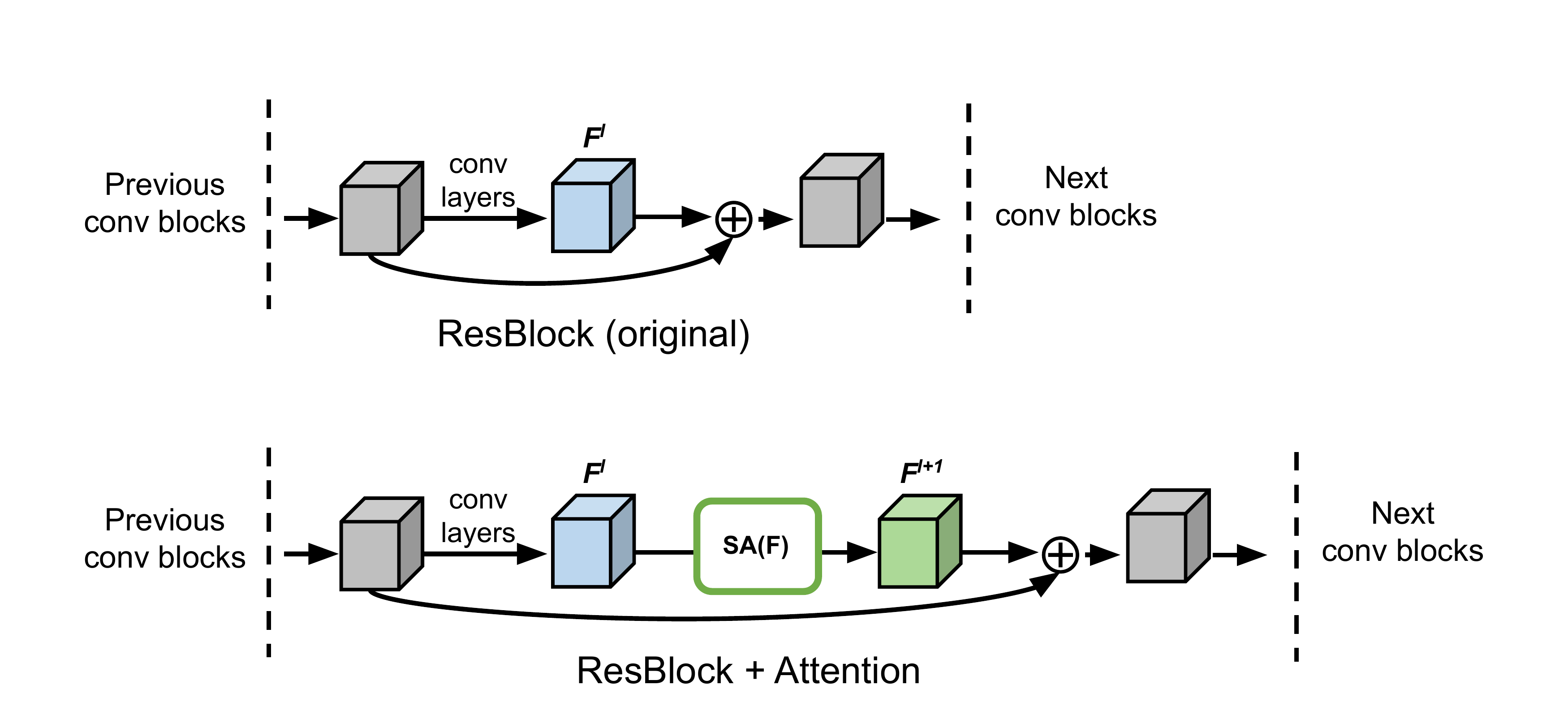}
    \caption{\textbf{Attention mechanism placement in ResBlocks.} Original ResBlock (top) and attention-augmented ResBlock (bottom) architectures. Placement consistency is maintained with a single modified ResBlock architecture.}
    \label{fig:att_placement}
\end{figure}

For the SE~\cite{hu2018squeeze} implementation, we used a third party PyTorch implementation\footnote{\url{https://github.com/StickCui/PyTorch-SE-ResNet}}, for CBAM~\cite{woo2018cbam}, we used the official PyTorch implementation\footnote{\url{https://github.com/Jongchan/attention-module}} and for GC~\cite{cao2019gcnet}, we used the official PyTorch implementation\footnote{\url{https://github.com/xvjiarui/GCNet}}. As illustrated in Fig.~\ref{fig:att_placement}, all the attention mechanisms (SE, CBAM, GC) are placed in each convolutional block in ResNet-18 right before the residual connection. The addition of attention within the ResBlock is the only difference between the standard and attention-augmented ResNet-18 model architectures.

\subsection{Implementation Details}\label{sec:implemen}

The models with and without the attention mechanisms are randomly initialized with \textit{He} initialization. The models have 2 output logits for each input and are trained end-to-end with the cross entropy loss. The stochastic gradient descent (SGD) optimizer is used across all experiments with a momentum of 0.9 and weight decay of $1e^{-4}$. For all experiments, we use a batch size of 64 and a initial learning rate 0.05. The learning rate is decayed by 0.1 at 30\%, 60\% and 90\% of the total epochs. For all datasets, we trained models for 100 epochs except the COVID-CT dataset which we trained across 500 epochs due to the limited sample count.

\section{Experimental Results}

In this section, we cover the experimental results derived through our validation of each model across the medical image datasets. The following text describes the quantitative and qualitative experiments conducted before presenting the results for each model and the corresponding datasets.

\noindent\paragraph{Quantitative}

In the quantitative experiments, we run each model variant across 3 individual tests and calculate the average area under the curve of receiver operating characteristic (AUC-ROC). AUC-ROC is a robust metric used to understand the performance of binary classification models. Averaging this metric across 3 tests can provide a more concrete reading which is not falsely elevated or deflated.

\noindent\paragraph{Qualitative}

For the qualitative experiments, we visualize Grad-CAM \cite{selvaraju2017grad} activation heat-maps across all model variants on different dataset image samples. We then conduct a anonymized user study in which different medical specialists are asked to interpret the visualizations and provide clinical insight onto which model is focusing on the most clinically relevant region(s) of the image. Questions inquired about for the visualizations are: (1). \textit{Which model focuses on the most important region?}, (2). \textit{Explain the answer in condensed clinical terminology?}. We perform this study for both the skin and CXR datasets with radiologists and dermatologists in order to gain clinical understanding.

\subsection{Skin Dataset}

\subsubsection{Quantitative}

The quantitative results for the skin cancer dataset are reported in Table~\ref{table:skin}. Row 1 depicts the baseline performance of ResNet-18 \cite{he2016deep}. ResNet-18 produced an average AUC-ROC of 93.28\% across 3 tests. The SE \cite{hu2018squeeze} modified ResNet-18 model (Row 2) received an average AUC-ROC of 95.06\% (+1.78 over ResNet-18). The CBAM \cite{woo2018cbam} modified ResNet-18 model (Row 3) received an average AUC-ROC score of 95.09\% (+1.81 over ResNet-18, +0.03 over ResNet-18 + SE). The GC \cite{cao2019gcnet} modified ResNet-18 model (Row 4) received an average AUC-ROC of 93.82\% (+0.54 over ResNet-18). The CBAM modified ResNet-18 model received the highest AUC-ROC score among the others. Additionally, through each individual test, CBAM and SE increase over baseline performance. Through this validation, we notice significant increases in AUC-ROC serving as a preliminary understanding of the benefits of attention.

\begin{table}
\centering
\resizebox{\columnwidth}{!}{%
\begin{tabular}{l|c|c|c|c}
Model  & Test 1           & Test 2           & Test 3           & Mean AUC-ROC      \\ 
\hline \hline
ResNet-18 \cite{he2016deep}         & 93.50\%          & 93.77\%          & 92.59\%          & 93.28\%          \\
\hline
ResNet-18 + SE \cite{hu2018squeeze}    & 95.20\% & \textbf{94.83}\% & 95.15\% & 95.06\% (+1.78) \\ 
ResNet-18 + CBAM \cite{woo2018cbam}  & \textbf{95.28}\%          & 94.73\%          & \textbf{95.26}\%          & \textbf{95.09}\% (+1.81)          \\ 
ResNet-18 + GC \cite{cao2019gcnet}     & 93.32\%          & 93.86\%          & 94.28\%          & 93.82\% (+0.54)          \\ 
\end{tabular}%
}
\caption{Quantitative Skin Cancer Dataset Results.}
\label{table:skin}
\end{table}

\subsubsection{Qualitative}

\begin{figure*}[]%
    \centering
    \subfloat[\centering Skin Dataset]{{\includegraphics[width=0.5\linewidth]{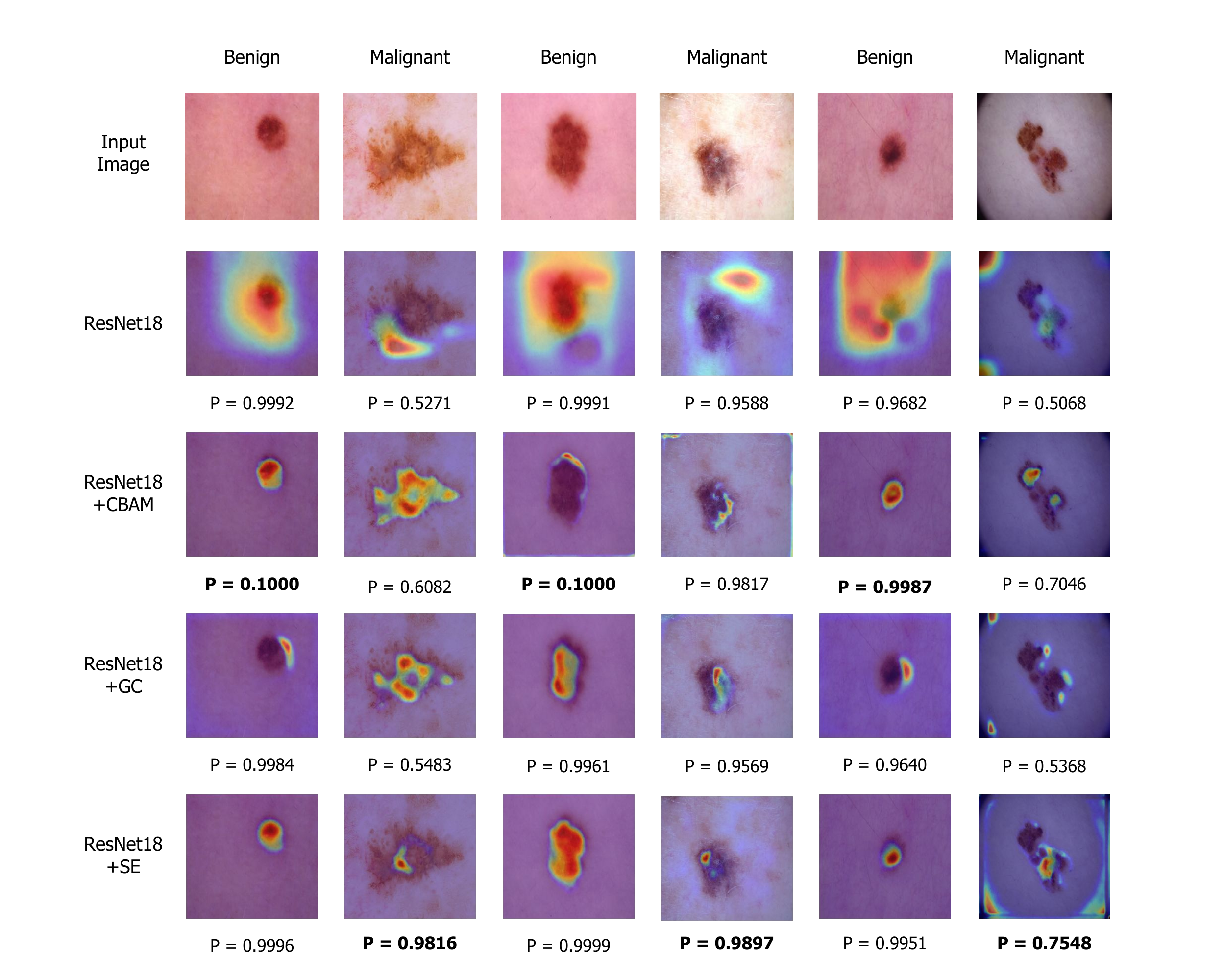} }}%
    \subfloat[\centering CXR Dataset]{{\includegraphics[width=0.5\linewidth]{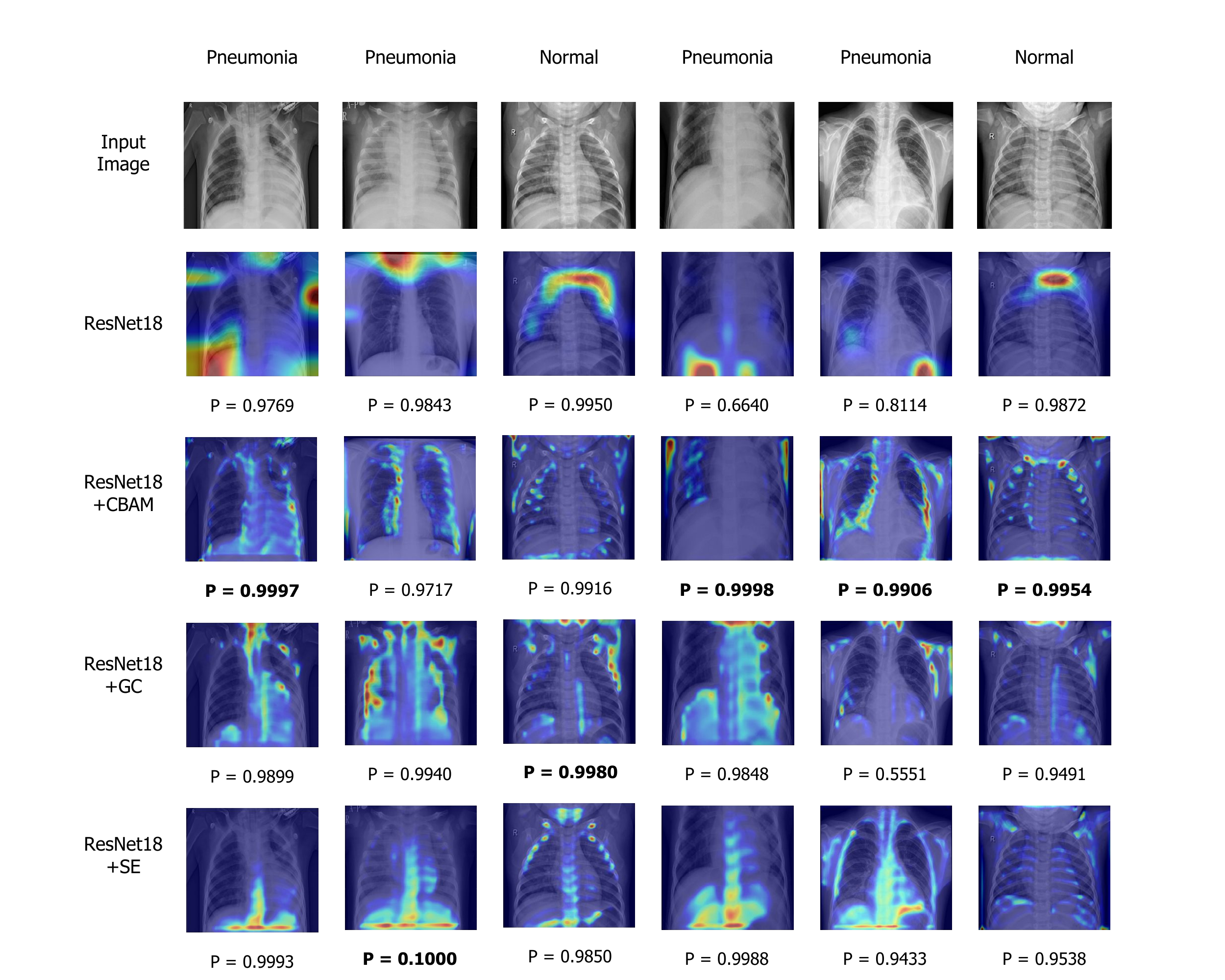} }}%
\caption{\textbf{Grad-CAM \cite{selvaraju2017grad} activation heat-map visualization from each model on the skin cancer and CXR datasets.} The visualization is generated from the last convolutional outputs. \textit{P} denotes the softmax classification percentage for the ground-truth prediction. Notice differences in heat-map and prediction percentages between each model.}
    \label{fig:skin_cancer_cxr}%
\end{figure*}

\begin{table}[]
\centering
\resizebox{\columnwidth}{!}{%
\begin{tabular}{|c|c|c|}
\hline
Column (\#) & Best Model & Description                                                                                                                                                                                                                                \\ \hline
1           & ResNet-18 + CBAM \cite{woo2018cbam}  & \begin{tabular}[c]{@{}c@{}}
"Malignant melanomas have asymmetrical shapes, irregular borders \\ and changes in color"\end{tabular}                                                                                                \\ \hline
2           & ResNet-18 + CBAM \cite{woo2018cbam}    & \begin{tabular}[c]{@{}c@{}}
"Malignant melanomas have asymmetrical shapes, irregular borders \\ 
and changes in color + the heat map for model 2 \\ 
for row 2 covers the greatest area of the actual lesion"\end{tabular}        \\ \hline
3           & ResNet-18 + SE \cite{hu2018squeeze}    & \begin{tabular}[c]{@{}c@{}}
"Malignant melanomas have asymmetrical shapes, irregular borders \\ 
and changes in color + the heat map for model 4 \\ 
for Column 3 covers the greatest area of the actual lesion."\end{tabular}    \\ \hline
4           & ResNet-18 + GC \cite{cao2019gcnet}    & \begin{tabular}[c]{@{}c@{}}
"Heat map focuses on irregularly raised area as well as \\ irregularly pigmented segments of lesion."\end{tabular}                                                                                    \\ \hline
5           & ResNet-18 + CBAM \cite{woo2018cbam}    & \begin{tabular}[c]{@{}c@{}}
"Malignant melanomas have asymmetrical shapes, irregular borders \\ and 
changes in color + the heat map for model 2 \\ 
for column 2 covers the greatest area of the actual lesion."\end{tabular} \\ \hline
6           & ResNet-18 + CBAM \cite{woo2018cbam}    & \begin{tabular}[c]{@{}c@{}}
"focuses on lesion. Model 4 looks great but focuses on border as well, \\
which is not as favorable. Not clear how much that affects result."\end{tabular}                                          \\ \hline
\end{tabular}%
}
\caption{\textbf{Skin dataset user-study survey results.} 'Best Model' is the model which focuses on the most clinically relevant region and 'Description' is a clinical explanation.}
\label{table:user_study_skin}
\end{table}

We use 6 images from the skin dataset and visualize ground-truth activation heat-maps using Grad-CAM \cite{selvaraju2017grad} for each model (baseline, SE model \cite{hu2018squeeze}, CBAM model \cite{woo2018cbam}, GC model \cite{cao2019gcnet}). We also report the softmax probability percentage for each prediction label. In a preliminary nature, it is evident that attention significantly changes the activation heat-map and increases prediction probability (Fig.~\ref{fig:skin_cancer_cxr}).


After generating the visualizations, we carried out the user study (Table \ref{table:user_study_skin}). A trained dermatologist interpreted the visualizations and answered the survey questions. The first piece of information is the "Best Model". This is what model the clinician feels has the best and greatest amount of activation over a clinically relevant region. The first observation that can be made is that all models selected across all columns were never the baseline (ResNet-18) and the majority of models selected was ResNet-18 + CBAM. The dermatologist described that "Malignant melanomas have asymmetrical shapes, irregular borders and changes in color" and that the models with attention "covers the greatest area of the actual lesion". Also mentioned was that models with attention focus on "irregularly raised area as well as irregularly pigmented segments of lesion". In summary, attention outperformed the baseline and ResNet-18 + CBAM had the highest AUC-ROC score and promising clinical user study results. The clinician also provided a summary in which they mentioned that self-attention "covered larger area of lesion/mole including border to normal skin" and was "not getting distracted by surrounding skin".

\subsection{CXR Dataset}

\subsubsection{Quantitative}

The quantitative results for the CXR dataset are reported in Table~\ref{table:cxr}. The standard ResNet-18 model \cite{he2016deep} (Row 1) received an average AUC-ROC of 97.27\% across 3 tests. The SE \cite{hu2018squeeze} modified ResNet-18 model (Row 2) received an average AUC-ROC of 99.13\% (+1.86 over ResNet-18). The CBAM \cite{woo2018cbam} modified model (Row 3) received an average AUC-ROC of 99.12\% (+1.85 over ResNet-18, -0.01 lower than ResNet-18 + SE). The GC \cite{cao2019gcnet} modified ResNet-18 model (Row 4) received an average AUC-ROC of 96.97\% (-0.4 lower than ResNet-18). CBAM and SE performed higher than the baseline. This validation showed that self-attention still increases performance however, GC performed lower (-0.4 lower than ResNet-18) and was to be further investigated in the following dataset experiments.

\begin{table}[]
\centering
\resizebox{\columnwidth}{!}{%
\begin{tabular}{l|c|c|c|c}
Model  & Test 1           & Test 2           & Test 3           & Mean AUC-ROC      \\ \hline\hline
ResNet-18 \cite{he2016deep}         & 98.57\%          & 97.85\%          & 95.40\%          & 97.27\%          \\ \hline
ResNet-18 + SE \cite{hu2018squeeze}    & 98.93\% & \textbf{99.15}\% & \textbf{99.30}\% & \textbf{99.13}\% (+1.86) \\ 
ResNet-18 + CBAM \cite{woo2018cbam}  & \textbf{99.22}\%          & 98.98\%          & 99.16\%          & 99.12\% (+1.85)          \\ 
ResNet-18 + GC \cite{cao2019gcnet}     & 98.16\%          & 96.43\%          & 96.01\%          & 96.87\% (-0.4)          \\ 
\end{tabular}%
}
\caption{Quantitative CXR Dataset Results.}
\label{table:cxr}
\end{table}

\begin{table}[]
\centering
\resizebox{\columnwidth}{!}{%
\begin{tabular}{|c|c|c|}
\hline
Column (\#) & Best Model & Description                                                                                                                                                                                                                                                                                        \\ \hline
1                                   & ResNet-18 + CBAM \cite{woo2018cbam}    & \begin{tabular}[c]{@{}c@{}}
"Model is focusing on lung, highlighting unilateral patchy areas of \\ 
consolidation, nodular opacities, bronchial wall thickening and \\ 
pleural effusions, not highlighting normal appearing lung parenchyma"\end{tabular}                                      \\ \hline
2                                   & ResNet-18 + CBAM \cite{woo2018cbam}    & \begin{tabular}[c]{@{}c@{}}
"Model is focusing on lung, highlighting unilateral patchy areas of \\ 
consolidation, nodular opacities, bronchial wall thickening and \\
pleural effusions, not highlighting normal appearing lung parenchyma. \\
Model does an excellent job here"\end{tabular} \\ \hline
3                                      & ResNet-18 + CBAM \cite{woo2018cbam}    & \begin{tabular}[c]{@{}c@{}}
"Model has the best ratio of highlighting \\ actual lung tissue vs. non-lung tissue"\end{tabular}                                                                                                                                                                    \\ \hline
4                                   & None       & \begin{tabular}[c]{@{}c@{}}
"The area of clinical is the left upper lobe and none of the models \\
do a great job of highlighting this are. Most models, \\
except Model 2 highlight a large portion non-lung tissue"\end{tabular}                                                                \\ \hline
5                                  & ResNet-18 + CBAM \cite{woo2018cbam}    & \begin{tabular}[c]{@{}c@{}}
"Does a fantastic job of focusing on right bronchial thickening \\
and patchy areas in right lower and middle lobes"\end{tabular}                                                                                                                                       \\ \hline
6                                      & ResNet-18 + CBAM \cite{woo2018cbam}    & \begin{tabular}[c]{@{}c@{}}
"Model 2 is doing a great job at picking the windows between the ribs to \\ 
identify the clear lung parenchyma. Still focusing a bit outside of \\
the lung but not more than the other models"\end{tabular}                                                       \\ \hline
\end{tabular}%
}
\caption{\textbf{CXR dataset user-study survey results.} 'Best Model' is the model which focuses on the most clinically relevant region and 'Description' is a clinical explanation.}
\label{table:user_study_cxr}
\end{table}

\subsubsection{Qualitative}

Similar to the skin dataset qualitative results, the qualitative results for the CXR dataset are shown in Fig.~\ref{fig:skin_cancer_cxr}. Again, as shown, attention increases classification probability and changes the heat-map in comparison to the ResNet-18 baseline. We can also see that the baseline is focusing on regions outside of the lung while attention starts to move the heat-map closer to the more important lung regions. The user-study results for the dataset are shown in Table \ref{table:user_study_cxr}.


\subsection{MRI Dataset}

The quantitative results for the MRI dataset are reported in Table~\ref{table:mri}. The standard ResNet-18 model \cite{he2016deep} received an average AUC-ROC of 93.68\%. The SE \cite{hu2018squeeze} modified ResNet-18 model has a average AUC-ROC of 96.03 (+2.35 over ResNet-18). The CBAM \cite{woo2018cbam} modified ResNet-18 model received an average AUC-ROC 96.77\% (+3.09 over ResNet-18, +0.74 over ResNet-18 + SE). The GC \cite{cao2019gcnet} modified ResNet-18 model received an average AUC-ROC of 91.25 (-2.43 lower than ResNet-18). Through this validation, we again notice increases in performance with SE and CBAM however, GC performed poorly in comparison to ResNet-18 proving that not all self-attention mechanisms will have a positive impact on medical classification tasks.

\begin{table}[]
\centering
\resizebox{\columnwidth}{!}{%
\begin{tabular}{l|c|c|c|c}
Model  & Test 1           & Test 2           & Test 3           & Mean AUC-ROC      \\ \hline\hline
ResNet-18 \cite{he2016deep}         & 93.15\%          & 92.49\%          & 95.37\%          & 93.68\%          \\ \hline
ResNet-18 + SE \cite{hu2018squeeze}    & \textbf{98.43}\% & 96.29\% & 93.34\% & 96.03\% (+2.35) \\ 
ResNet-18 + CBAM \cite{woo2018cbam}  & 95.74\%          & \textbf{97.30}\%          & \textbf{97.25}\%          & \textbf{96.77}\% (+3.09)          \\ 
ResNet-18 + GC \cite{cao2019gcnet}     & 93.01\%          & 87.44\%          & 93.28\%          & 91.25\% (-2.43)          \\ 
\end{tabular}%
}
\caption{Quantitative MRI Dataset Results.}
\label{table:mri}
\end{table}

\subsection{CT Dataset}

The quantitative results for the CT dataset are shown in Table~\ref{table:ct}. The standard ResNet-18 model \cite{he2016deep} received an average AUC-ROC of 84.80\% (Row 1). The SE \cite{hu2018squeeze} modified ResNet-18 model (Row 2) received an average AUC-ROC of 88.14\% (+3.34 over ResNet-18). The CBAM \cite{woo2018cbam} modified ResNet-18 (Row 3) received an average AUC-ROC of 91.02\% (+6.22 over ResNet-18, +2.88 over ResNet-18 + SE). The GC \cite{cao2019gcnet} modified ResNet-18 (Row 4) received average AUC-ROC of 84.23\% (-0.57 lower than ResNet-18).

\begin{table}[]
\centering
\resizebox{\columnwidth}{!}{%
\begin{tabular}{l|c|c|c|c}
Model  & Test 1           & Test 2           & Test 3           & Mean AUC-ROC      \\ \hline\hline
ResNet-18 \cite{he2016deep}         & 87.79\%          & 87.61\%          & 78.99\%          & 84.80\%          \\ \hline
ResNet-18 + SE \cite{hu2018squeeze}    & 85.43\% & 90.86\% & 88.13\% & 88.14\% (+3.34) \\
ResNet-18 + CBAM \cite{woo2018cbam}  & \textbf{90.37}\%          & \textbf{93.25}\%          & \textbf{89.45}\%          & \textbf{91.02}\% (+6.22)          \\
ResNet-18 + GC \cite{cao2019gcnet}     & 84.68\%          & 82.36\%          & 85.66\%          & 84.23\% (-0.57)          \\ 
\end{tabular}%
}
\caption{Quantitative CT Dataset Results.}
\label{table:ct}
\end{table}











\section{Conclusion}

In this paper, we evaluate various self-attention mechanisms within medical computer vision systems. Self-attention enables standard CNN models to focus more on semantically important or aggregated relevant content within features. The use of attention improved the AUC-ROC for medical vision task accuracy on dermatologic melanoma images, CXR images, brain MRI images and COVID-19 CT scans. Clinical user-study survey reviews conferred greater clinical agreement with feature focus of self-attention mechanism heat-maps. Further validation with other datasets and attention is required to further validate the improved accuracy trend observed with attention. 


{\small
\bibliographystyle{ieee}
\bibliography{egbib}

\begin{thebibliography}{10}\itemsep=-1pt

\bibitem{isicarchive}
Isic archive.

\bibitem{albawi2017understanding}
S.~Albawi, T.~A. Mohammed, and S.~Al-Zawi.
\newblock Understanding of a convolutional neural network.
\newblock In {\em 2017 International Conference on Engineering and Technology
  (ICET)}, pages 1--6. Ieee, 2017.

\bibitem{allport1989visual}
A.~Allport.
\newblock Visual attention.
\newblock 1989.

\bibitem{bertram2016eye}
R.~Bertram, J.~Kaakinen, F.~Bensch, L.~Helle, E.~Lantto, P.~Niemi, and
  N.~Lundbom.
\newblock Eye movements of radiologists reflect expertise in ct study
  interpretation: A potential tool to measure resident development.
\newblock {\em Radiology}, 281(3):805--815, 2016.

\bibitem{bhandary2020deep}
A.~Bhandary, G.~A. Prabhu, V.~Rajinikanth, K.~P. Thanaraj, S.~C. Satapathy,
  D.~E. Robbins, C.~Shasky, Y.-D. Zhang, J.~M.~R. Tavares, and N.~S.~M. Raja.
\newblock Deep-learning framework to detect lung abnormality--a study with
  chest x-ray and lung ct scan images.
\newblock {\em Pattern Recognition Letters}, 129:271--278, 2020.

\bibitem{bhattacharyya2011brain}
D.~Bhattacharyya and T.-h. Kim.
\newblock Brain tumor detection using mri image analysis.
\newblock In {\em International Conference on Ubiquitous Computing and
  Multimedia Applications}, pages 307--314. Springer, 2011.

\bibitem{cao2019gcnet}
Y.~Cao, J.~Xu, S.~Lin, F.~Wei, and H.~Hu.
\newblock Gcnet: Non-local networks meet squeeze-excitation networks and
  beyond.
\newblock In {\em Proceedings of the IEEE/CVF International Conference on
  Computer Vision Workshops}, pages 0--0, 2019.

\bibitem{chun2001visual}
M.~M. Chun and J.~M. Wolfe.
\newblock Visual attention.
\newblock {\em Blackwell handbook of perception}, 272310, 2001.

\bibitem{attention2}
G.~Corbetta~M., Shulman.
\newblock Control of goal-directed and stimulus-driven attention in the brain.
\newblock {\em Nature Reviews Neuroscience}, 3:201–215, 2002.

\bibitem{cozzi2020chest}
D.~Cozzi, M.~Albanesi, E.~Cavigli, C.~Moroni, A.~Bindi, S.~Luvar{\`a},
  S.~Lucarini, S.~Busoni, L.~N. Mazzoni, and V.~Miele.
\newblock Chest x-ray in new coronavirus disease 2019 (covid-19) infection:
  findings and correlation with clinical outcome.
\newblock {\em La radiologia medica}, 125:730--737, 2020.

\bibitem{esteva2021deep}
A.~Esteva, K.~Chou, S.~Yeung, N.~Naik, A.~Madani, A.~Mottaghi, Y.~Liu,
  E.~Topol, J.~Dean, and R.~Socher.
\newblock Deep learning-enabled medical computer vision.
\newblock {\em NPJ digital medicine}, 4(1):1--9, 2021.

\bibitem{grewal2018radnet}
M.~Grewal, M.~M. Srivastava, P.~Kumar, and S.~Varadarajan.
\newblock Radnet: Radiologist level accuracy using deep learning for hemorrhage
  detection in ct scans.
\newblock In {\em 2018 IEEE 15th International Symposium on Biomedical Imaging
  (ISBI 2018)}, pages 281--284. IEEE, 2018.

\bibitem{he2016deep}
K.~He, X.~Zhang, S.~Ren, and J.~Sun.
\newblock Deep residual learning for image recognition.
\newblock In {\em Proceedings of the IEEE conference on computer vision and
  pattern recognition}, pages 770--778, 2016.

\bibitem{he2020sample}
X.~He, X.~Yang, S.~Zhang, J.~Zhao, Y.~Zhang, E.~Xing, and P.~Xie.
\newblock Sample-efficient deep learning for covid-19 diagnosis based on ct
  scans.
\newblock {\em medrxiv}, 2020.

\bibitem{hu2018squeeze}
J.~Hu, L.~Shen, and G.~Sun.
\newblock Squeeze-and-excitation networks.
\newblock In {\em Proceedings of the IEEE conference on computer vision and
  pattern recognition}, pages 7132--7141, 2018.

\bibitem{huang2019ccnet}
Z.~Huang, X.~Wang, L.~Huang, C.~Huang, Y.~Wei, and W.~Liu.
\newblock Ccnet: Criss-cross attention for semantic segmentation.
\newblock In {\em Proceedings of the IEEE/CVF International Conference on
  Computer Vision}, pages 603--612, 2019.

\bibitem{attention1}
L.~Itti, C.~Koch, and E.~Niebur.
\newblock A model of saliency-based visual attention for rapid scene analysis.
\newblock {\em IEEE Transactions on Pattern Analysis and Machine Intelligence},
  20(11):1254--1259, 1998.

\bibitem{jerant2000early}
A.~F. Jerant, J.~T. Johnson, C.~D. Sheridan, and T.~J. Caffrey.
\newblock Early detection and treatment of skin cancer.
\newblock {\em American family physician}, 62(2):357--368, 2000.

\bibitem{ker2017deep}
J.~Ker, L.~Wang, J.~Rao, and T.~Lim.
\newblock Deep learning applications in medical image analysis.
\newblock {\em Ieee Access}, 6:9375--9389, 2017.

\bibitem{kermany2018labeled}
D.~Kermany, K.~Zhang, M.~Goldbaum, et~al.
\newblock Labeled optical coherence tomography (oct) and chest x-ray images for
  classification.
\newblock {\em Mendeley data}, 2(2), 2018.

\bibitem{korolev2017residual}
S.~Korolev, A.~Safiullin, M.~Belyaev, and Y.~Dodonova.
\newblock Residual and plain convolutional neural networks for 3d brain mri
  classification.
\newblock In {\em 2017 IEEE 14th international symposium on biomedical imaging
  (ISBI 2017)}, pages 835--838. IEEE, 2017.

\bibitem{lakshmanaprabu2019optimal}
S.~Lakshmanaprabu, S.~N. Mohanty, K.~Shankar, N.~Arunkumar, and G.~Ramirez.
\newblock Optimal deep learning model for classification of lung cancer on ct
  images.
\newblock {\em Future Generation Computer Systems}, 92:374--382, 2019.

\bibitem{lee2019srm}
H.~Lee, H.-E. Kim, and H.~Nam.
\newblock Srm: A style-based recalibration module for convolutional neural
  networks.
\newblock In {\em ICCV}, 2019.

\bibitem{lee2017deep}
J.-G. Lee, S.~Jun, Y.-W. Cho, H.~Lee, G.~B. Kim, J.~B. Seo, and N.~Kim.
\newblock Deep learning in medical imaging: general overview.
\newblock {\em Korean journal of radiology}, 18(4):570--584, 2017.

\bibitem{litjens2017survey}
G.~Litjens, T.~Kooi, B.~E. Bejnordi, A.~A.~A. Setio, F.~Ciompi, M.~Ghafoorian,
  J.~A. Van Der~Laak, B.~Van~Ginneken, and C.~I. S{\'a}nchez.
\newblock A survey on deep learning in medical image analysis.
\newblock {\em Medical image analysis}, 42:60--88, 2017.

\bibitem{liu2018applications}
J.~Liu, Y.~Pan, M.~Li, Z.~Chen, L.~Tang, C.~Lu, and J.~Wang.
\newblock Applications of deep learning to mri images: A survey.
\newblock {\em Big Data Mining and Analytics}, 1(1):1--18, 2018.

\bibitem{rajpurkar2017chexnet}
P.~Rajpurkar, J.~Irvin, K.~Zhu, B.~Yang, H.~Mehta, T.~Duan, D.~Ding, A.~Bagul,
  C.~Langlotz, K.~Shpanskaya, et~al.
\newblock Chexnet: Radiologist-level pneumonia detection on chest x-rays with
  deep learning.
\newblock {\em arXiv preprint arXiv:1711.05225}, 2017.

\bibitem{attention3}
R.~A. Rensink.
\newblock The dynamic representation of scenes.
\newblock {\em Visual cognition}, 7(1-3):17--42, 2000.

\bibitem{sartaj_2020}
Sartaj.
\newblock Brain tumor classification (mri), May 2020.

\bibitem{selvaraju2017grad}
R.~R. Selvaraju, M.~Cogswell, A.~Das, R.~Vedantam, D.~Parikh, and D.~Batra.
\newblock Grad-cam: Visual explanations from deep networks via gradient-based
  localization.
\newblock In {\em Proceedings of the IEEE international conference on computer
  vision}, pages 618--626, 2017.

\bibitem{shen2017deep}
D.~Shen, G.~Wu, and H.-I. Suk.
\newblock Deep learning in medical image analysis.
\newblock {\em Annual review of biomedical engineering}, 19:221--248, 2017.

\bibitem{wang2017residual}
F.~Wang, M.~Jiang, C.~Qian, S.~Yang, C.~Li, H.~Zhang, X.~Wang, and X.~Tang.
\newblock Residual attention network for image classification.
\newblock In {\em Proceedings of the IEEE conference on computer vision and
  pattern recognition}, pages 3156--3164, 2017.

\bibitem{wang2018non}
X.~Wang, R.~Girshick, A.~Gupta, and K.~He.
\newblock Non-local neural networks.
\newblock In {\em Proceedings of the IEEE conference on computer vision and
  pattern recognition}, pages 7794--7803, 2018.

\bibitem{woo2018cbam}
S.~Woo, J.~Park, J.-Y. Lee, and I.~S. Kweon.
\newblock Cbam: Convolutional block attention module.
\newblock In {\em Proceedings of the European conference on computer vision
  (ECCV)}, pages 3--19, 2018.

\bibitem{yang2020covid}
X.~Yang, X.~He, J.~Zhao, Y.~Zhang, S.~Zhang, and P.~Xie.
\newblock Covid-ct-dataset: a ct scan dataset about covid-19.
\newblock {\em arXiv preprint arXiv:2003.13865}, 2020.

\end{thebibliography}
}

\end{document}